\documentclass[final]{pasj00}

\Received{$\langle$reception date$\rangle$}
\Accepted{$\langle$acception date$\rangle$}
\Published{$\langle$publication date$\rangle$}
\SetRunningHead{Astronomical Society of Japan}{Usage of \texttt{pasj00.cls}}

\usepackage{times}

\begin{document}

\title{Appearance of Jet-Driving Poynting Flux in Hot, Tenuous Accretion Disks \\ 
Threaded by an Ordered Magnetic Field}

\author{Osamu \textsc{Kaburaki},
}
\affil{Otomaru-machi 249, Hakusan-shi, Ishikawa 924-0826
}
\email{okab@amail.plala.or.jp}


\maketitle

\begin{abstract}

In a series of our previous works, a model of radiatively inefficient accretion flows (RIAFs) in a global 
magnetic field (so called resistive RIAF model) has proved its ability to account for many physical 
processes taking place in such accretion flows as realized in the nuclei of the galaxies believed to 
be accreting at a very small fraction of each Eddinton accretion rate. 
Within the present status of this model, however, the model cannot describe the launch of 
a self-confined bipolar jet from the vicinity of disk's inner edge, although it allows the existence 
of a thermal wind widely distributed over the disk surfaces. 
This is because the electric field (and hence the Poynting flux) vanishes everywhere in the disk, 
whereas such a jet in a globally ordered magnetic field is most likely to be accelerated 
electrodynamically. 
We show in the present paper that this defect can be overcome naturally if we reformulate the problem 
so as to admit a quasi-stationary change of the magnetic field (and hence the appearance of a 
non-irrotational electric field), and also restore all the terms of order 
$\epsilon\equiv (v_r/v_{\varphi})^2\lesssim 1$ (where $v_r$ and $v_{\varphi}$ denote radial and 
azimuthal components, respectively, of the fluid velocity) which have been neglected altogether in 
the previous treatments. The restored effects are the inertial and magnetic draggings 
on the infalling matter. As an illustrative example, a model solution which is correct up to 
${\cal O}(\epsilon)$ is derived under a set of plausible restrictions. The new solution predicts 
the appearance of a localized Poynting flux in a region near the disk inner edge, suggesting strongly 
that a jet is launched from this region. Another interesting prediction is the appearance of a rapid 
change of the magnetic field also localized to this region.   

\end{abstract}


\section{Introduction}

Well organized astrophysical jets are often observed around various central objects, e.g., both 
supermassive and stellar mass black holes, neutron stars, and even young forming stars. Among them, 
Galactic black-hole X-ray binaries (BHXBs) have served as good laboratories for stellar mass black holes 
because of their closeness to the Earth. 
It has been confirmed mainly from their radio and X-ray observations that the presence or absence of 
a jet, and if present, the nature of the jet from such a BHXB has close correlation with the state of 
associated accretion disk (e.g., \cite{FBG04}). As often argued by many authors (e.g., \cite{F10}), 
such a correlation between jets and accretion disks in BHXBs may be considered to apply also to their 
scale-up version, the active galactic nuclei (AGN). This conspicuous correlation between the jet state 
and associated accretion disk strongly suggests that the main source of the jet-driving power is in 
the accretion power rather than in the spin energy of the central black holes. It is, therefore, natural 
to demand that any accretion disk (or flow) theory that ever concerns with the inner regions 
where a jet might be launched should process a due ability to describe this circumstances. 

According to the terminology by \citet{NM08} \footnote{however, see also footnote 2}, inefficiently 
radiating accretion flows (RIAFs) form the low mass-accretion rate branch (see, e.g., \cite{FKR02}; 
\cite{KFM08}) of the series of states called advection dominated accretion flows (ADAFs). 
The other separated branch is that for slim disks. 
RIAFs are specified by a radiation cooling time longer than the accretion time, while the slim disks 
have a diffusion time for photons longer than the accretion time. In both cases, therefore, the radiation 
does not have sufficient time to carry away significant part of energy from accreting matter, and hence 
main part of energy may be advected down the flow in the form of thermal energy. 
The critical accretion rate which divides the two branches seems to be 
$\dot{m}_{\rm crit}\simeq0.01\sim0.1$, when it is normalized to the Eddington rate. 

Currently, the RIAFs are considered to be the most plausible accretion states that may be accompanied 
with well defined jets. Indeed, overall structure of the broad-band spectral energy distributions 
(SEDs) from the nuclei of radio-loud galaxies, which are the most typical central engines of well-developed 
jets, have been adequately reproduced by the RIAF models (e.g.,\cite{Nry02}; \cite{Yun07}). 
We know also that strong jets appear in the low/hard state of the BHXBs, and the state has been 
understood as the RIAF dominated one (e.g., \cite{FBG04}; \cite{Nry05}). 
A subclass of the radio-loud AGN, the low excitation (or ionization) radio galaxies, has been 
recognized as the counterpart of the low/hard state in BHXBs (\cite{Bld10}) and they also can be 
understood as in RIAF dominated states.

The basic ideas of RIAF models have been developed by many researchers (e.g., \cite{Ich77}; 
\cite{RBBF82}; \authorcite{NY94}\yearcite{NY94, NY95}; \cite{Abr95}). One of the variations of RIAFs, 
the convection dominated accretion flows (CDAFs), have also been proposed in this context 
(\cite{IA00}; \cite{NIA00}, \cite{QG00}), in which convective transport of energy is dominant in the 
accreting matter instead of adiabatic infall in the case of original RIAFs. 
A number of review papers on RIAFs and related flows have been published in the literature (e.g., 
\cite{Lst99}; \cite{Nry02}; \cite{NM08}). 

The ability of RIAFs to drive associated outflows has been pointed out from the viewpoint of their 
positive Bernoulli sums (i.e., sum of potential energy, kinetic energy and enthalpy) by \authorcite{NY94} 
(\yearcite{NY94, NY95}). As a few authors have suggested (\cite{Nak98}; \cite{ALI00}), however, 
the problem seemed to be not so straightforward and need much careful considerations. 
Afterwards, several analytical works (\cite{BB99}; \cite{Bck00}; \cite{TD00}) have indicated that 
what is actually needed is to clarify the process of energy transfer from the infalling part (i.e., 
accreting matter) with negative Bernoulli sum to the outflowing part (i.e., wind and/or jet) with 
positive sum. Anyway, those outflows modeled explicitly in the above mentioned works seem to be widely 
distributed winds rather than much concentrated jets.  

While the gas dynamical version of the RIAF models mentioned above is widely known, it is also 
widely accepted that the presence of an ordered magnetic field in the nuclear region is quite essential 
to the jet formation (\cite{Liv99}; \cite{Pll04}; \cite{DalP05}). Therefore, a model of the RIAF 
penetrated by an ordered magnetic field seems to be the most promising version to account for the jet 
launching. Indeed, many numerical works has been performed to confirm this point (e.g., \cite{MKU01}; 
\cite{DalP05}). 
Nevertheless it is strongly desirable to construct an analytic model, in order to separately discuss 
various aspects of physical processes taking place in such accretion flows and outflows. 

One of such models for RIAF in a global magnetic field has been developed by the present author 
in the last decade (\authorcite{Kab00}\yearcite{Kab00, Kab01, Kab07}) and applied successfully to the 
nuclei of some typical dim galaxies (\cite{KKY00}; \cite{YKK02}) and radio-loud galaxies (\cite{KNTW10}). 
Hereafter we call it the `resistive' RIAF model, in order to distinguish it from the well known `viscous' 
RIAF model in which the presence of a global magnetic field is neglected. 

In the former model, the resistive heating (probably of anomalous type) due to the electric current 
caused in an accretion disk plays a dominant role in dissipating the gravitational energy, instead of 
the viscous heating in the latter model. Although some types of small-scale, turbulent magnetic fields 
are included in the viscous RIAF model through a parameterization, it is a great advantage of the 
resistive RIAF model that the model can self-consistently specify the deformation by the flow, of an 
externally given ordered magnetic field. 
Another noticeable difference between the two versions of the models is in their disk thickness: 
resistive RIAFs are geometrically thin while viscous RIAFs are thick, although both have a similar 
viral-like high temperature. This is because, in the former case, the toroidal magnetic field 
generated by the action of a rotating accretion disk develops mainly outside the disk and compresses 
the disk plasma toward its midplane (\cite{Kab07}). 

Within the resistive RIAF model, it has been shown clearly that a thermal wind can be generated according 
to the state of transfer of thermal energy through the plasma (\cite{Kab01}). 
The direction (i.e., upward or downward) and strength of a wind are specified by a parameter $n$. 
When $n=0$ the flow is adiabatic (i.e., there is no energy transfer through the accreting plasma) 
and there is no wind. When $n$ is positive/negative (corresponding to the additional 
input/output of thermal energy by some process like conduction, convection or radiation) 
\footnote{Generally speaking, a non-radiating accretion flow is not necessarily be dominated by advection. 
It may be dominated by convection and/or conduction. Therefore, the name ADAFs should be reserved strictly 
for the special case of general RIAFs (i.e., no-radiating flows), in which there is no exchange of energy 
among the fluid elements (i.e., adiabatic). In that case, we can use the term RIAFs for both optically 
thin and thick branches, but have to invent a new name to call optically thin branch distinctively. }
there is upward/downward wind from the surfaces of the disk. The larger value of 
$\vert n\vert$ means that the larger fraction of matter concerns with a wind. 

Unfortunately, however, the model in the current status cannot describe the launch of any electrodynamic 
jet which is expected to occur near the inner edge of an accretion disk. This is because the electric 
field predicted by this model vanishes identically in the disk, and hence, also does the Poynting flux 
that is necessary for the jet driving.  In the present paper, we overcome this defect by improving the 
accuracy of the model. 

First in section 2, the framework of the resistive RAIF model is reconsidered to accommodate the problem 
of electrodynamic jet launching and reformulated as a quasi-stationary problem. Then in section 3, each 
physical quantity is anticipated to be written as a multiple of two parts, which are the functions only 
of radius and of co-latitude, respectively, within the geometrically thin-disk approximation. Such a 
separation of variables is completed in section 4 within a certain degree of approximation, and a set of 
ordinary differential equations is derived for functions of radius only. The method adopted in 
sections 3 and 4 is essentially the same as in the current version of the resistive RIAF model. Section 5 
is devoted to the discussion of the previously obtained solution, the vanishing Poynting-flux (VPF) 
solution, from our new standpoint. The derivation of a model solution, the non-VPF solution, 
to the newly formulated problem is given in section 6. In section 7, the Poynting flux and the Bernoulli 
sum in this solution are calculated. The time variability of the magnetic field generated by our model 
accretion disks is evaluated also in this section. We summarize the results in the final section. 

\section{Quasi-Stationary Problem}

As will be shown in section \ref{chap:VPF}, the vanishing of the electric field (and hence, of Poynting 
flux) everywhere in the accretion disk in our previous works is closely related to the assumption of 
strict stationarity. Therefore, we need a careful reformulation of the problem which we really have 
to deal with. 

For this purpose, we first introduce two characteristic time scales associated with the accretion 
processes in an ordered magnetic filed. 
One is the accretion time defined by 
 \begin{equation}
   \tau_{\rm acc} \equiv \frac{L}{v_r}, 
 \label{eqn:tacc}
 \end{equation}
where $L$ is a representative length scale and $v_r$ is the radial velocity component (in spherical 
polar coordinates, $r$, $\theta$, $\varphi$), respectively, of the accretion flow. 
The other is the electrodynamic time defined by 
 \begin{equation}
   \tau_{\rm ed} \equiv \frac{LB}{cE}, 
 \label{eqn:ted}
 \end{equation}
which can be derived from Faraday's law (equation (\ref{eqn:Farad}) below). In this paper, we use CGS 
Gaussian units, and $c$, $E$ and $B$ here denote the light velocity, characteristic sizes of the 
electric and magnetic fields, respectively.  

Quasi-stationary accretion flows in an ordered magnetic field are defined as the flows that satisfy 
the following requirements: 
\begin{enumerate}
 \item an accretion flow obeys the condition, 
    $\epsilon\equiv (v_r/v_{\varphi})^2 \lesssim 1$,
 \item the fluid motion is stationary, 
 \item electrodynamic processes are controlled by the accretion processes, and hence, 
    $\tau_{\rm ed}=\tau_{\rm acc}$. 
\end{enumerate}

In the close vicinity of a central black hole, the infall velocity may become $v_r \sim v_{\varphi}$ 
or even $v_r > v_{\varphi}$ owing to the inward attraction of the gravity and effective extraction 
of angular momentum by the magnetic field. Such a flow is usually called plunging flow, distinguished 
from the accretion flows which satisfy $(v_r/v_{\varphi})^2 \lesssim 1$. In this sense, the inner edge 
of an accretion disk (or flow) should be identified with the place where $v_r \sim v_{\varphi}$ 
(or $\epsilon \sim 1$). In our current version of the resistive RIAF model, the requirement 1 
is surely satisfied (\cite{Kab00}, \yearcite{Kab01}).

Note that, in the above definition, the stationarity is required only for the fluid motion, but not 
for the electromagnetic field. The time scale for the variation of the electromagnetic field is 
determined automatically from the set of relevant equations so as to satisfy the item 3 of the above 
requirements. In this sense, the elctrodynamic processes subordinate the accretion processes. 
Requiring the strict stationarity also for the electric field would be too restrictive and constrains 
the field to be irrotational (i.e., electrostatic). In this step, the most vital aspect of the accretion 
process may have been lost. 

The requirement 3 above yields 
 \begin{equation}
  E \sim \frac{v_r B}{c}. 
 \end{equation}
Hereafter, we assume that the accretion  flows are non-relativistic in the sense $\delta\equiv 
(v_{\varphi}/c)^2\lesssim (v_{\rm K}/c)^2\ll 1$,  where $v_{\rm K}$ denotes the Kepler velocity. 
This assumption may be justified, because the inner boundaries of resistive RIAFs have turned out to be 
located at about a hundred Schwartzschild radii \citep{KKY00,YKK02,KNTW10}, where even the Kepler 
velocity is non-relativistic. Then we have
 \begin{displaymath}
   \left(\frac{E}{B}\right)^2 \sim \left(\frac{v_r}{c}\right)^2 
     = \left(\frac{v_{\varphi}}{c}\right)^2\left(\frac{v_r}{v_{\varphi}}\right)^2 
   = \epsilon\delta \lesssim \delta \ll 1. 
 \end{displaymath}
In a resistive RIAF whose inner edge is located at such radii, any general relativistic effect 
may also be neglected. 

Thus, the set of non-relativistic magnetohydrodynamic (MHD) equations applied to the quasi-stationary 
problem becomes as follows: 
\begin{itemize}
 \item leading equations
  \begin{equation} 
        {\bf \nabla}\cdot\mbox{\boldmath $B$} = 0, 
   \label{eqn:Fcont}
  \end{equation}
  \begin{equation}
       \mbox{\boldmath $j$} = \frac{c}{4\pi}\ {\bf \nabla}\times\mbox{\boldmath $B$}, 
   \label{eqn:Amp}
  \end{equation}
  \begin{equation}
    \mbox{\boldmath $E$} = \frac{\mbox{\boldmath $j$}}{\sigma} 
       - \frac{1}{c}\ \mbox{\boldmath $v$}\ \times\mbox{\boldmath $B$}, 
   \label{eqn:Ohm}
  \end{equation}
  \begin{equation} 
           {\bf \nabla}\cdot(\rho\mbox{\boldmath $v$}) = 0, 
   \label{eqn:Mcont}
  \end{equation}
  \begin{equation}
           (\mbox{\boldmath $v$}\cdot{\bf \nabla})\mbox{\boldmath $v$} = 
           - \frac{1}{\rho}\ {\bf \nabla}p 
           - \frac{1}{4\pi\rho}\ [\mbox{\boldmath $B$}\times({\bf \nabla}\times\mbox{\boldmath $B$})] 
           + \mbox{\boldmath $g$}, 
  \label{eqn:EOM}
  \end{equation}
  \begin{equation}
        p = K \rho T,
  \label{eqn:EOS}
  \end{equation}

 \item subsidiary equations

  \begin{equation} 
        \frac{1}{c}\frac{\partial \mbox{\boldmath $B$}}{\partial t} 
            = - {\bf \nabla}\times \mbox{\boldmath $E$}, 
  \label{eqn:Farad}
  \end{equation}
  \begin{equation}
        q = \frac{1}{4\pi}\ {\bf \nabla}\cdot\mbox{\boldmath $E$}.
  \label{eqn:chg}
  \end{equation}
\end{itemize}
The pressure $p$ is assumed to satisfy the ideal gas law, in which $T$ and $\rho$ are the temperature 
and density of the fluid, respectively, and $K$ is the gas constant divided by the mean molecular weight. 
The charge and current densities are denoted by $q$ and $\mbox{\boldmath $j$}$, respectively. 
The gravitational force per unit mass, $\mbox{\boldmath $g$}$, is simply the Newton gravity which has 
only the $r$-component, 
\begin{equation}
  g_r = -\frac{GM}{r^2}, 
\end{equation} 
where $G$ is the gravitational constant and $M$ is the mass of a central black hole. 

Equation (\ref{eqn:Amp}) is Amp\`{e}re's law which neglects the displacement current in the corresponding 
Maxell's equation. As stated above, the stationarity of the electric field is not automatically 
guaranteed in the quasi-stationary problem. Nevertheless, the neglection of the displacement current 
can be justified since 
\begin{displaymath}
  \left|\frac{1}{c}\frac{\partial\mbox{\boldmath $E$}}{\partial t}\right| \biggm/ 
  \vert{\bf \nabla}\times\mbox{\boldmath $B$}\vert 
  \sim\frac{E}{c\tau_{\rm ed}} \biggm/ \frac{B}{L} 
  \sim \left(\frac{E}{B}\right)^2 \lesssim \delta \ll 1. 
\end{displaymath}
Equation (\ref{eqn:Ohm}) is Ohm's law in which $\sigma$ represents the electric conductivity and 
is assumed to be constant, for simplicity. 
Fluid motion is governed by equation of motion (EOM), (\ref{eqn:EOM}), and mass continuity, 
(\ref{eqn:Mcont}). The time derivative terms in both equations have been dropped according to the 
assumption of quasi-stationarity. 
Also in equation (\ref{eqn:EOM}), the electric force in the Lorentz formula has been neglected since 
\begin{displaymath}
  \vert q\mbox{\boldmath $E$} \vert 
  \biggm/ \left|\frac{1}{c}\mbox{\boldmath $j$}\times\mbox{\boldmath $B$}\right| 
  \sim\left(\frac{E}{B}\right)^2 \lesssim \delta \ll 1, 
\end{displaymath}
where $q\sim E/L$ and $j\sim cB/L$ has been used. Thus, the above set of MHD equations are correct 
in every region of the accretion flow where $\epsilon \lesssim 1$ is satisfied. 

We can evaluate the charge density and the time rate of magnetic filed from the above subsidiary 
equations, after the electromagnetic field has been determined. Local budget of the electromagnetic 
energy is described by the Poynting theorem, 
\begin{equation}
  \frac{\partial u}{\partial t} = -{\bf \nabla}\cdot\mbox{\boldmath $P$}
   -\mbox{\boldmath $E$}\cdot\mbox{\boldmath $j$},
\end{equation}
which can be derived from Maxwell's equations. Here, $u$ and $\mbox{\boldmath $P$}$ are the 
electromagnetic energy density and the Poynting flux, respectively, defined by the relations 
\begin{equation}
  u \equiv \frac{1}{8\pi}(E^2+B^2) \approx \frac{B^2}{8\pi}, 
  \quad \mbox{\boldmath $P$} \equiv \frac{c}{4\pi}\mbox{\boldmath $E$}\times\mbox{\boldmath $B$}. 
\end{equation}

It should be noted here that the energy equation for a fluid is lacking in the above set of MHD 
equations. Usually, this equation is replaced by the assumption of polytropic law for the fluid, 
since the inclusion of the former makes the problem hopelessly difficult to treat analytically. 
Also in the resistive RIAF model, a parameter analogous to the polytropic index is introduced, 
which is related to the affairs of energy transport within it. This point will turn out explicit in 
section \ref{chap:VPF}. 

\section{Axisymmetric, Geometrically Thin Disks}

In the resistive RIAF model, the accretion disk is assumed to be axisymmetric and geometrically thin. 
The former assumption is only for simplicity but the plausibility of the latter assumption is 
confirmed within the framework of this model. Indeed, the model suggests (\cite{Kab00}, 
\yearcite{Kab01}, \yearcite{Kab07}) that a strong azimuthal component is generated from a seed magnetic 
field and develops within the upper and lower halves of such a jet bubble as observed in soft X-rays 
(e.g., \cite{WSY06}, \cite{Frm07}, \cite{Krf07}). The disk is, therefore, confined to a thin structure 
located at the equatorial plane of a bipolar jet system, owing to its magnetic pressure.
When we denote the half opening-angle of an accretion disk by $\Delta (\ll 1)$, the upper and lower 
surfaces of the disk is defined as $\theta=\pi/2\pm\Delta$, respectively. 

As in our previous works, we adopt the method of approximate separation of variables in order to reach 
an analytic solution to the quasi-stationary accretion problem, based on the fact that the disk is 
geometrically thin. The smallness parameter here is $\Delta$. Reflecting this fact, it is convenient 
to introduce a normalized colatitude, $\eta \equiv (\theta-\pi/2)/\Delta$. 
On the other hand, the radius from the center is normalized to the radius of disk's inner edge 
$r_{\rm in}$, i.e., $\xi\equiv r/r_{\rm in}$. Note that the definition of $\xi$ in this paper is 
different from those in our previous papers: e.g., $\xi$ in \citet{Kab00} corresponds to $\eta$ here, 
and that in \authorcite{Kab01} (\yearcite{Kab01}, \yearcite{Kab07}) implies $\xi\equiv r/r_{\rm out}$, 
where $r_{\rm out}$ is the outer edge radius. 

Another fact which is useful for simplifying the problem is a large deformation of the seed field. 
In a resistive RIAF, the ordered magnetic field penetrating the accretion disk is largely amplified 
from an interstellar seed field by the action of both infall and rotational motions in the disk. Since 
the deformed part of the field, $\mbox{\boldmath$b$}$, is much larger than the original seed field, 
$\mbox{\boldmath$B$}_0$, except in the region very close to the outer edge, we use the approximation 
\begin{equation}
  \mbox{\boldmath$B$} = \mbox{\boldmath$B$}_0 + \mbox{\boldmath$b$} \approx \mbox{\boldmath$b$}. 
\end{equation}
Therefore in the set of MHD equations introduced in section 3, every $\mbox{\boldmath$B$}$ can be 
replaced by $\mbox{\boldmath$b$}$. After this replacement, the component expressions of the MHD set 
in spherical polar coordinates become to the forms cited in Appendix. 

The seed field $\mbox{\boldmath$B$}_0$ here is considered as a uniform vector vertical to an 
accretion disk, idealizing a large scale interstellar magnetic field. The deformed part of the field is 
maintained by the electric current generated in the disk, whose poloidal component has been suggested to 
close after circulating through the boundary of a large scale X-ray cavity and the polar regions 
\citep{Kab07}. The resulting field outside the disk is a helical field confined to this X-ray cavity. 
Such physical considerations are reflected as boundary conditions to $\mbox{\boldmath$b$}(\xi, \eta)$ 
in selecting the functional forms given below. Namely, the poloidal components of the deformed field 
$\mbox{\boldmath$b$}$ should vanish at large $\eta$, while its toroidal component can remain 
there. The boundary condition at the disk outer edge is $\vert\mbox{\boldmath$b$}\vert \rightarrow 
\vert\mbox{\boldmath$B$}_0\vert$. We simply regard that their equality as a definition of the 
outer edge. 

According to our previous works, we expect separations of variables for every relevant physical 
quantity in the following form: 
\begin{equation}
  b_r(\xi,\ \eta) =\ \tilde{b}_r(\xi)\ \mbox{sech}^2\eta\ \tanh\eta,  
 \label{eqn:br}
\end{equation}
\begin{equation}
  b_{\theta}(\xi,\ \eta) =\ \Delta\tilde{b}_{\theta}(\xi)\ \mbox{sech}^2\eta, 
 \label{eqn:bth}
\end{equation}
  \begin{equation}
     b_{\varphi}(\xi,\ \eta) =-\tilde{b}_{\varphi}(\xi)\tanh\eta, 
 \label{eqn:bph}
  \end{equation}
  \begin{equation}
     j_r(\xi,\ \eta) =-\Delta^{-1}\tilde{j}_r(\xi)\ \mbox{sech}^2\eta, 
 \label{eqn:jr}
  \end{equation}
  \begin{equation}
     j_{\theta}(\xi,\ \eta) = \tilde{j}_{\theta}(\xi)\ \tanh\eta,
 \label{eqn:jth}
  \end{equation}
  \begin{equation}
     j_{\varphi}(\xi,\ \eta) =-\Delta^{-1}\tilde{j}_{\varphi}(\xi)\ \mbox{sech}^2\eta 
      \ (\mbox{sech}^2\eta - 2\tanh\eta), 
 \label{eqn:jph}
  \end{equation}
\begin{equation}
  E_r(\xi,\ \eta) = \Delta\tilde{E}_r(\xi)\ \mbox{sech}^2\eta, 
 \label{eqn:Er}
\end{equation}
\begin{equation}
  E_{\theta}(\xi,\ \eta) = -\tilde{E}_{\theta}(\xi)\ \mbox{sech}^2\eta\ \tanh\eta,  
 \label{eqn:Eth}
\end{equation}
\begin{equation}
  E_{\varphi}(\xi,\ \eta) = -\Delta\tilde{E}_{\varphi}(\xi)\ \mbox{sech}^2\eta, 
 \label{eqn:Eph}
\end{equation}
  \begin{equation}
     v_r(\xi,\ \eta) = -\tilde{v}_r(\xi)\ (\mbox{sech}^2\eta - 2\tanh^2\eta), 
 \label{eqn:vr}
  \end{equation}
  \begin{equation}
     v_{\theta}(\xi,\ \eta) = \Delta\tilde{v}_{\theta}(\xi)\ \tanh\eta,
 \label{eqn:vth}
  \end{equation}
  \begin{equation}
     v_{\varphi}(\xi,\ \eta) = \tilde{v}_{\varphi}(\xi), 
 \label{eqn:vph}
  \end{equation}
  \begin{equation}
     p(\xi,\ \eta) =\ \tilde{p}(\xi)\ \mbox{sech}^2\eta, 
 \label{eqn:p}
  \end{equation}
  \begin{equation}
     \rho(\xi,\ \eta) =\ \tilde{\rho}(\xi)\ \mbox{sech}^2\eta, 
 \label{eqn:rho}
  \end{equation}
  \begin{equation}
   T(\xi,\ \eta) = \tilde{T}(\xi). 
 \label{eqn:T}
  \end{equation}

Each quantity has been written as a multiplication of its radial and angular parts, which are respectively 
the functions of $\xi$ and $\eta$ only. The radial-part functions are distinguished from 
the corresponding total functions by tildes on them. The relative orders of magnitude among the above 
quantities are indicated explicitly in terms of $\Delta$ included in the above expressions. Therefore, 
all functions with tilde on them may be considered as of order unity with respect to $\Delta$. It may 
be interesting to see that among others $v_{\varphi}$ and $T$ are assumed to be independent of $\eta$. 

\section{Equations for Radial-Part Functions}

Given the variable separated expressions, (\ref{eqn:br}) through (\ref{eqn:T}), the next task is to 
actually perform the separation of variables on each equation cited in Appendix 1 in order to obtain 
the corresponding equation containing only radial-part functions. Of course, however, we cannot expect 
the exact separation of variables among so many equations, most of which are so complex in 
$\eta$-dependences. Instead, we have to be satisfied with a certain degree of approximation. The status 
we keep in this paper (and has been kept in our related previous papers) is as follows: i) in equations 
as many as possible, the separation should be completed within the leading order terms in the smallness 
parameter $\Delta$, and ii) if the separation is not performed even among the leading order terms in an 
equation, it should be done at least near the midplane of the accretion disk (i.e., in the limit of 
$\eta\rightarrow 0$ neglecting the terms of order higher than $\mbox{tanh}^2 \eta$). 

Of course, the separation restricted to the limit of $\eta\rightarrow 0$ is invalid near and outside 
the surfaces of an accretion disk (i.e., $\eta\rightarrow\infty$). In this sense, the above approximation 
scheme means that we focus our primary attention on the main body of the disk, where accreting matter is 
concentrated and expected to be governing the physical processes.
Another strategy to analytically deal with such geometrically thin disks may be the averaging over 
the disk height (e.g., \cite{SS73}). We believe, however, that the above introduced method of 
approximate separation of variables has many advantages over that of height averaging in extracting 
much physical information contained in the problem. 

Noting that $b_{r}\sim{\cal O}(1)$ and $b_{\theta}\sim{\cal O}(\Delta)$, we obtain 
\begin{equation}
    \frac{1}{r^2}\frac{\partial}{\partial r}(r^2 b_r) 
    + \frac{1}{r}\ \frac{\partial b_{\theta}}{\partial \theta} = 0 
  \end{equation}
as the leading order approximation of equation (\ref{eqn:fcont}). 
The term $(1/r) \partial{b_{\theta}}/\partial{\theta}$, which results from the second derivative 
in equation (\ref{eqn:fcont}), is a quantity of order ${\cal O}(1)$, 
since $\partial/\partial\theta = \Delta^{-1}\partial/\partial\eta \sim {\cal O}(\Delta^{-1})$. 
The other term $(\cot\theta/r)b_{\theta}$, resulting also from the second derivative in equation 
(\ref{eqn:fcont}), has been dropped because it is a quantity of ${\cal O}(\Delta^2)$ in the disk where 
$\theta-\pi/2 < \Delta$. Then, the separation can be performed perfectly in this leading order equation, 
yielding a useful expression 
 \begin{equation}
  \frac{\tilde{b}_{\theta}}{\tilde{b}_r} = \frac{r}{2}\frac{d}{dr}\ln(r^2\tilde{b}_r) 
  = \frac{\xi}{2}\frac{d}{d\xi}\ln(\xi^2\tilde{b}_r). 
 \label{eqn:fcontS}
 \end{equation}

Similarly, equation of mass continuity becomes 
  \begin{equation}
    \frac{1}{r^2}\frac{\partial}{\partial r}(r^2\rho v_r) 
    + \frac{1}{r}\ \frac{\partial}{\partial \theta}(\rho v_{\theta}) = 0,  
  \end{equation}
in the leading order approximation, and, further substituted the expressions (\ref{eqn:vr}), 
(\ref{eqn:vth}) and (\ref{eqn:rho}), reduces to 
 \begin{equation}
  \frac{\tilde{v}_{\theta}}{\tilde{v}_r} = r\frac{d}{dr}\ln(r^2\tilde{\rho}\tilde{v}_r) 
  = \xi\frac{d}{d\xi}\ln(\xi^2\tilde{\rho}\tilde{v}_r). 
  \label{eqn:mcontS}
 \end{equation}
Also in this case, the separation of variables is exact among the leading order terms. An example of 
exactly separated case to all orders in $\Delta$ is the equation of state, which reduces to 
  \begin{equation}
     \tilde{p} = K\tilde{\rho}\tilde{T}. 
  \label{eqn:eqsS}
  \end{equation}

The $\eta$-dependences of the current density components, (\ref{eqn:jr}) $\sim$ (\ref{eqn:jph}), have 
been determined from the forms of the magnetic field components, (\ref{eqn:br}) $\sim$ (\ref{eqn:bph}), 
through the leading oder versions of Amp\`{e}re's law, 
  \begin{equation}
    \frac{4\pi}{c} j_r 
       = \frac{1}{r}\ \frac{\partial b_{\varphi}}{\partial\theta}, 
  \end{equation}
  \begin{equation}
    \frac{4\pi}{c} j_{\theta}
       = - \frac{1}{r}\ \frac{\partial}{\partial r} (r b_{\varphi}), 
  \end{equation}
  \begin{equation}
    \frac{4\pi}{c} j_{\varphi}
       = - \frac{1}{r}\ \frac{\partial b_r}{\partial\theta}. 
  \end{equation}
This process yields also  
  \begin{equation}
    \tilde{j}_r = \frac{c\tilde{b}_{\varphi}}{4\pi r}, 
    \quad \tilde{j}_{\theta} = \frac{c}{4\pi r}\ \frac{d}{dr}(r\tilde{b}_{\varphi}),
    \quad \tilde{j}_{\varphi} = \frac{c\tilde{b}_r}{4\pi r}, 
  \label{eqn:currS}
  \end{equation}
as a result of variable separations. 

The leading order versions of the component expressions of Ohm's law are 
  \begin{equation}
    E_r = \frac{j_r}{\sigma} - \frac{1}{c}(v_{\theta}b_{\varphi} - v_{\varphi}b_{\theta}), 
  \end{equation}
  \begin{equation}
    E_{\theta} = - \frac{1}{c}(v_{\varphi}b_r - v_r b_{\varphi}), 
  \end{equation}
  \begin{equation}
    E_{\varphi} = \frac{j_{\varphi}}{\sigma} - \frac{1}{c}(v_r b_{\theta} - v_{\theta}b_r), 
  \end{equation}
Both $r$- and $\varphi$-components contain full terms, while the term $j_{\theta}/\sigma$ has been 
omitted in the $\theta$-component equation. This is because we regard $\Delta^2\sigma \sim {\cal O}(1)$ 
(consistent with the fact that plasma is a good conductor), according to our previous works. For these 
equations, the separation of variables cannot be performed even among the leading order terms so that 
we must be satisfied with the separation in the limit of $\eta\rightarrow 0$ 
(i.e., $\mbox{sech}^2\eta\rightarrow 1$, and $\tanh^2\eta\rightarrow 0$). 
The results are  
  \begin{equation}
    \tilde{E}_r = \frac{1}{c}\ \tilde{v}_{\varphi}\tilde{b}_{\theta} 
        - \frac{\tilde{j}_r}{\Delta^2\sigma}, 
  \label{eqn:ErS}
  \end{equation}
  \begin{equation}
    \tilde{E}_{\theta} = - \frac{1}{c}\ (\tilde{v}_{\varphi}\tilde{b}_r 
       - \tilde{v}_r \tilde{b}_{\varphi}), 
  \label{eqn:EthS}
  \end{equation}
  \begin{equation}
    \tilde{E}_{\varphi} = - \frac{1}{c}\ \tilde{v}_r\tilde{b}_{\theta} 
       + \frac{\tilde{j}_{\varphi}}{\Delta^2\sigma}. 
  \label{eqn:EphS}
  \end{equation}

Although equation of motion is the most complicated one in our set of equations, it can be treated 
also within the above introduced scheme of approximate variable separations. In the leading order 
approximation in $\Delta$, their three components become 
  \begin{eqnarray}
      \lefteqn{ \left( v_r \frac{\partial}{\partial r} 
         + \frac{v_{\theta}}{r} \frac{\partial}{\partial \theta} \right)v_r 
         - \frac{v_{\varphi}^2}{r} }\nonumber \\
       & & \quad = -\frac{1}{\rho}\ \frac{\partial p}{\partial r} -\frac{GM}{r^2} 
           - \frac{1}{4\pi\rho r} \left[ b_{\varphi} \frac{\partial}{\partial r}(rb_{\varphi})
           - b_{\theta}\frac{\partial b_r}{\partial \theta} \right],
  \end{eqnarray}
  \begin{equation}
     \frac{\partial}{\partial\theta}\left\{ p+\frac{1}{8\pi}(b_r^2 + b_{\varphi}^2) \right\} =0, 
  \end{equation}
  \begin{eqnarray}
      \lefteqn{ \left( v_r \frac{\partial}{\partial r} 
         + \frac{v_{\theta}}{r} \frac{\partial}{\partial \theta} \right)v_{\varphi} 
         + \frac{v_r v_{\varphi}}{r} }\nonumber \\
      & & \quad = \frac{1}{4\pi\rho r} \left[ b_r \frac{\partial}{\partial r}(r b_{\varphi}) 
       + b_{\theta} \frac{\partial b_{\varphi}}{\partial \theta} \right].
  \end{eqnarray}

Among them, the $\theta$-component has been so simplified that it can be integrated 
to yield 
\begin{equation}
   p + \frac{1}{8\pi}(b_r^2 + b_{\varphi}^2) = \mbox{const.}, 
\end{equation}
where `const.' on the right-hand side means that it is independent of $\theta$ (i.e., it may be a 
function of $\xi$ only). When we denote this function as $\tilde{p}(\xi)$, the separation of 
variables can be performed in the limit of $\eta\rightarrow 0$, and we have 
\begin{equation}
   \tilde{p} = \frac{1}{8\pi}(\tilde{b}_r^2 + \tilde{b}_{\varphi}^2). 
  \label{eqn:EOMthS}
\end{equation}
Roughly speaking, this means a balance between the gas pressure on the midplane of a disk and the 
magnetic pressure due to both $\varphi$- and $r$-components at the disk surfaces. 
The other two equations are separated also in the same limit: 
  \begin{equation}
      \tilde{v}_r \frac{d\tilde{v}_r}{dr} 
         - \frac{\tilde{b}_r\tilde{b}_{\theta}}{4\pi\tilde{\rho} r} 
       = \frac{\tilde{v}_{\varphi}^2}{r}-\frac{1}{\tilde{\rho}}
         \ \frac{d\tilde{p}}{dr} -\frac{GM}{r^2}, 
  \label{eqn:EOMrS}
  \end{equation}
\begin{equation}
    \tilde{v}_r \frac{d}{dr}(r\tilde{v}_{\varphi}) 
      - \frac{\tilde{b}_{\theta}\tilde{b}_{\varphi}}{4\pi\tilde{\rho}} = 0. 
  \label{eqn:EOMphS}
\end{equation}

Finally, we go on to the subsidiary equations. The leading order approximations of Faraday's law 
can be written as, 
  \begin{equation}
   \frac{1}{c}\ \frac{\partial b_r}{\partial t} 
   = -\frac{1}{r}\ \frac{\partial E_{\varphi}}{\partial\theta}, 
  \end{equation}
  \begin{equation}
   \frac{1}{c}\ \frac{\partial b_{\theta}}{\partial t} 
   = \frac{1}{r}\ \frac{\partial}{\partial r}(r E_{\varphi}), 
  \end{equation}
  \begin{equation}
   \frac{1}{c}\ \frac{\partial b_{\varphi}}{\partial t} 
   = -\frac{1}{r}\ \left\{ \frac{\partial}{\partial r}(rE_{\theta}) 
      -\frac{\partial E_r}{\partial\theta} \right\}. 
  \end{equation}
The latter two equations are the same as their full versions, i.e., no term has been neglected. 
The former two can be separated for any value of $\eta$, resulting in 
  \begin{equation}
   \frac{1}{c}\ \frac{\partial \tilde{b}_r}{\partial t} 
   = -2\frac{\tilde{E}_{\varphi}}{r}, 
  \label{eqn:FdrS}
  \end{equation}
  \begin{equation}
   \frac{1}{c}\ \frac{\partial \tilde{b}_{\theta}}{\partial t} 
   = -\frac{1}{r}\ \frac{d}{dr}(r\tilde{E}_{\varphi}), 
  \label{eqn:FdthS}
  \end{equation}
and the latter one results in 
  \begin{equation}
   \frac{1}{c}\ \frac{\partial \tilde{b}_{\varphi}}{\partial t} 
   = -\frac{1}{r}\ \left\{ \frac{d}{dr}(r\tilde{E}_{\theta}) 
      - 2 \tilde{E}_r \right\}, 
  \label{eqn:FdphS}
  \end{equation}
 when separated in the limit of $\eta\rightarrow 0$. 

The functional form of the charge density is determined from the right-hand side of equation 
(\ref{eqn:chgA}). Keeping only the leading order term on the right-hand side, we have 
  \begin{equation}
   q = \frac{1}{4\pi r}\frac{\partial E_{\theta}}{\partial\theta}, 
  \label{eqn:chgS}
  \end{equation}
 and further substituting the expression (\ref{eqn:Eth}), 
\begin{eqnarray}
   q(\xi,\ \eta) &=& -\Delta^{-1}\tilde{q}(\xi)\ \mbox{sech}^2\eta\ 
         (\mbox{sech}^2\eta-2\tanh^2\eta), \nonumber \\
   \tilde{q}(\xi) &=& \frac{\tilde{E}_{\theta}(\xi)}{4\pi r}. 
  \label{eqn:FchgS}
\end{eqnarray}



\section{VPF Solution, Revisited} \label{chap:VPF}

Under the assumption of strict stationarity, together with a few conditions to select a RIAF state, 
we have obtained an analytic solution to the set of MHD equations (\authorcite{Kab00} \yearcite{Kab00}, 
\yearcite{Kab01}). In that solution, however, the electric field identically vanishes in the accretion 
flow. Therefore, there is no Poynting flux to drive such outside systems as bipolar jets often observed 
to be emanating from the nuclear regions. The whole electric power generated by plasma motion in the 
magnetic field is merely dissipated locally in the accretion disk, as Ohmic dissipation. Hereafter, 
we call this solution the vanishing Poynting-flux (VPF) solution. 

In this section, we first observe that the vanishing of the electric field is closely related to 
the assumption of strict stationarity, and then rederive the VPF solution starting from the vanishment 
of the electric filed. This section may also serve the reader as a brief introduction to the 
current status of our resistive RIAF solution.

It is convenient to rewrite the electric field components, (\ref{eqn:ErS}) 
$\sim$ (\ref{eqn:EphS}), as 
  \begin{equation}
    \tilde{E}_r = -\frac{1}{c}\ \tilde{v}_r\tilde{b}_{\theta}  
        \left( \frac{\Re_t}{\Re_p} - \frac{\tilde{v}_{\varphi}}{\tilde{v}_r} \right), 
  \label{eqn:ErM}
  \end{equation}
  \begin{equation}
    \tilde{E}_{\theta} = - \frac{1}{c}\ \tilde{v}_r\tilde{b}_r 
        \left( \Re_t - \frac{\tilde{v}_{\varphi}}{\tilde{v}_r} \right), 
  \label{eqn:EthM}
  \end{equation}
  \begin{equation}
    \tilde{E}_{\varphi} = - \frac{1}{c}\ \tilde{v}_r\tilde{b}_{\theta} \left( 1 - \Re_p^{-1} \right), 
  \label{eqn:EphM}
  \end{equation}
where the poloidal and toroidal magnetic Reynolds numbers have been introduced by the relations 
  \begin{equation}
   \Re_p \equiv \frac{4\pi\Delta^2\sigma r\tilde{v}_r}{c^2}\ 
      \frac{\tilde{b}_{\theta}}{\tilde{b}_r}, \qquad 
   \Re_t \equiv \frac{\tilde{b}_{\varphi}}{\tilde{b}_r}, 
  \label{eqn:defRe}
  \end{equation}
respectively. The former of the above definitions is introduced anew in the present paper. 

The requirement of strict stationarity means the vanishing of the left-hand side of equation 
(\ref{eqn:Farad}). In our case, this yields  
\begin{equation}
  \tilde{E}_{\varphi} = 0, \qquad \frac{d}{dr}(r\tilde{E}_{\varphi}) = 0, 
  \qquad \tilde{E}_r =\frac{1}{2}\frac{d}{dr}(r\tilde{E}_{\theta}), 
\end{equation}
from its variable separate counterparts, (\ref{eqn:FdrS}), (\ref{eqn:FdthS}) and (\ref{eqn:FdphS}). 
The second of the above equations is always satisfied if the first one holds. 
The implications of the third equation are classified into two cases according to the value of 
$\tilde{E}_{\theta}$. When $\tilde{E}_{\theta} = 0$, the equation requires also $\tilde{E}_r = 0$. 
This is the case of vanishing electric field, $\tilde{\mbox{\boldmath $E$}}=0$. 
When $\tilde{E}_{\theta}\neq 0$, then dividing both sides of the equation by $\tilde{E}_{\theta}$ and 
substituting equation (\ref{eqn:fcontS}), we have $\tilde{E}_{\theta}/r\tilde{b}_r 
= \tilde{E}_r/r\tilde{b}_{\theta} = \mbox{const.}$. 
As will be shown below, the former case leads to the VPF solution derived in our previous papers. 
However, we don't know at present whether there is a consistent set of solution in the latter case. 

Our re-derivation of the VPF solution begins with $\tilde{E}_{\varphi}=0$, which leads to $\Re_p=1$ or 
\begin{equation}
   \tilde{v}_r = \frac{c^2}{4\pi\Delta^2\sigma r}\ \frac{\tilde{b}_r}{\tilde{b}_{\theta}}.  
 \label{eqn:vrM}
\end{equation}
The result means that the infall velocity of the accreting matter is determined not from EOM but 
from Ohm's law as a diffusion velocity through the ordered poloidal magnetic field. When $\Re_p=1$, 
both conditions $\tilde{E}_r = 0$ and $\tilde{E}_{\theta} = 0$ yield the same result, 
\begin{equation}
  \Re_t = \frac{\tilde{v}_{\varphi}}{\tilde{v}_r}\ (= \epsilon^{-1/2}). 
  \label{eqn:Reyt}
\end{equation}
As stated in section 2, the accretion flow is specified by the condition $\epsilon\lesssim 1$ so that 
its inner boundary should be identified with the place where $\epsilon=1$, i.e., $\Re_t=1$ in this case. 

Originally, the VPF solution has been found in the effort searching for a state with quasi-Keplerian 
rotation. The requirements for a quasi-Keplerian rotation are the following: 
\begin{enumerate}
  \item slow infall, i.e., $\epsilon\equiv(\tilde{v}_r/\tilde{v}_{\varphi})^2 \ll 1$, so that equation 
   (\ref{eqn:EOMrS}) reduces to 
   \begin{equation}
      \frac{\tilde{v}_{\varphi}^2}{r}-\frac{1}{\tilde{\rho}}
            \ \frac{d\tilde{p}}{dr} -\frac{GM}{r^2} = {\cal O}(\epsilon)\rightarrow 0, 
   \label{eqn:EOMrM}
   \end{equation}
  \item gas temperature with viral form,  
   \begin{equation}
     \alpha \equiv -\frac{1}{\tilde{\rho}}\ \frac{d\tilde{p}}{dr}\Bigm/\frac{GM}{r^2}
     = \mbox{const.}, \quad (0<\alpha<1). 
   \label{eqn:alpha}
   \end{equation}
\end{enumerate}

The first requirement is well justified in an accretion disk not so close to its inner edge, and when 
the terms of order $\epsilon$ are completely neglected, the rotational velocity is determined by the 
gravity partly cancelled by the contribution from the pressure gradient term. The second requirement 
assures that the temperature is of a virial type (i.e., $\tilde{T}\propto 1/r$) as long as the pressure 
$\tilde{p}$ is a power law function of $r$. This can be confirmed by rewriting it in the form 
$\alpha=K\tilde{T}(-d\ln\tilde{p}/dr)/(GM/r^2)$. 
From these conditions, we actually have a quasi-Keplerian rotation, 
\begin{equation}
  \tilde{v}_{\varphi} = (1-\alpha)^{1/2}V_{\rm K}, 
\end{equation}
where $V_{\rm K} = (GM/r)^{1/2}$ denotes the Kepler velocity. 

The determination of other quantities is straightforward, if we introduce the functional form 
\begin{equation}
  \tilde{b}_r(\xi) = B_{\rm in}\ \xi^{-(3/2-n)}, 
  \label{eqn:brV}
\end{equation}
where $B_{\rm in}\equiv B_0\ \xi_{\rm out}^{3/2-n}$. The form of $B_{\rm in}$ guarantees that the size 
of the deformed field $\mbox{\boldmath $b$}$ should be comparable to that of the external field $B_0$ 
at the outer edge of an accretion disk, $\xi_{\rm out}$. 
The expectation of this kind of power law form for $\tilde{b}_r$ comes from the fact that $\tilde{T}$ 
and $\tilde{v}_{\varphi}$ have already been introduced as power law functions. 

The appearance of the unspecified power law index $-(3/2-n)$ reflects the lack of energy equation 
from our basic equations, and the separation of the factor 3/2 is only for convenience. In this 
sense, the power law index $n$ may be regarded as a free parameter analogous to the polytropic 
index. Indeed, it has been shown that its values reflect the thermal energy budget of a fluid element 
in an accretion disk (see, section 7 of \cite{Kab01}). Positive and negative values of $n$ 
correspond to a gain and a loss, respectively. The energy transfer may be caused by conduction, 
convection or radiation. Specifically, $n=0$ corresponds to the case of adiabatic flows, i.e., there is 
no exchange of energy among the neighboring fluid elements. Thus, the role of the parameter $n$ is quite 
analogous to that of the polytropic index. 

Substituting the expression (\ref{eqn:brV}) into equation (\ref{eqn:fcontS}), we have 
$\tilde{b}_{\theta}/\tilde{b}_r=(2n+1)/4$ and hence 
\begin{equation}
  \tilde{b}_{\theta}(\xi) = \frac{2n+1}{4}B_{\rm in}\ \xi^{-(3/2-n)}. 
\end{equation}
The result can be used to derive 
\begin{equation}
  \tilde{v}_r(\xi) = v_{\rm in}\ \xi^{-1},  
  \quad v_{\rm in} \equiv \frac{c^2}{(2n+1)\pi\Delta^2\sigma r_{\rm in}}, 
  \label{eqn:vrm}
\end{equation}
directly from equation (\ref{eqn:vrM}), and further combined with equation (\ref{eqn:Reyt}), 
\begin{equation}
  \Re_t(\xi) = \Re_{\rm in}\xi^{1/2}, \quad 
      \Re_{\rm in}\equiv\frac{v_{\rm in}}{(1-\alpha)^{1/2}V_{\rm K,in}}, 
\end{equation}
where $V_{\rm K,in} \equiv (GM/r_{\rm in})^{1/2}$. Simultaneously, we find $\Re_{\rm in}=1$ also from 
the expression (\ref{eqn:Reyt}), since by its definition $\Re_t(1)=1$ should hold at the inner edge 
($\xi=1$) of a disk. Therefore, the final expression of the toroidal magnetic Reynolds number becomes 
\begin{equation}
  \Re_t(\xi) = \xi^{1/2}. 
  \label{eqn:Ret} 
\end{equation}
Then, the result 
\begin{equation}
  \tilde{b}_{\varphi}(\xi) = B_{\rm in}\ \xi^{-(1-n)} 
 \label{eqn:bphV}
\end{equation}
follows from the definition of $\Re_t$. 

The components of the current density calculated from the set of equations (\ref{eqn:currS}) are 
  \begin{equation}
    \tilde{j}_r(\xi) = \frac{cB_{\rm in}}{4\pi r_{\rm in}}\ \xi^{-(2-n)}, 
  \end{equation}
  \begin{equation}
    \tilde{j}_{\theta}(\xi) = n\left(\frac{cB_{\rm in}}{4\pi r_{\rm in}}\right)\ \xi^{-(2-n)}, 
   \label{eqn:jthVPF}
  \end{equation}
  \begin{equation}
    \tilde{j}_{\varphi}(\xi) = \frac{cB_{\rm in}}{4\pi r_{\rm in}}\ \xi^{-(5/2-n)}. 
  \end{equation}
Although we can derive the explicit expression for the inner edge radius, $r_{\rm in}$, from the above 
definitions of $v_{\rm in}$ and $\Re_{\rm in}$, it would be practically useless owing to large 
ambiguities contained in the size of the electrical conductivity $\sigma$. Rather, it may be useful 
to evaluate effective values of the conductivity, regarding $r_{\rm in}$ as a parameter which can 
be inferred from, e.g., spectral fittings (e.g., \cite{KKY00}; \cite{YKK02}). 

Neglecting the term of order $\epsilon$ as in equation (\ref{eqn:EOMrM}), the gas pressure 
(\ref{eqn:EOMthS}) can be approximated like 
\begin{equation}
  \tilde{p} = \frac{\tilde{b}_{\varphi}^2}{8\pi}\left(1+\Re_t^{-2}\right) 
            = \frac{\tilde{b}_{\varphi}^2}{8\pi}\left(1+\epsilon\right)
           \rightarrow \frac{\tilde{b}_{\varphi}^2}{8\pi}. 
\end{equation}
Then, substituting equation (\ref{eqn:bphV}), we have  
\begin{equation}
  \tilde{p}(\xi) = \frac{B_{\rm in}^2}{8\pi}\ \xi^{-2(1-n)}. 
\end{equation}
The gas density is calculated from equation (\ref{eqn:EOMphS}) to give 
\begin{equation}
  \tilde{\rho}(\xi) = \rho_{\rm in}\ \xi^{-(1-2n)}, \quad 
   \rho_{\rm in} \equiv \frac{2(n+1)}{8\pi(1-\alpha)^{1/2}}\ \frac{B_{\rm in}^2}{v_{\rm in}V_{\rm K,in}}. 
\end{equation}
These results enables us to determine the value of $\alpha$ from the definition (\ref{eqn:alpha}) as 
\begin{equation}
  \alpha = \frac{2}{3}(1-n), 
  \label{eqn:alphaF}
\end{equation}
where the relations 
\begin{eqnarray}
  - \frac{1}{\tilde{\rho}}\ \frac{d\tilde{p}}{dr} 
   &=& \frac{2(1-n)(1-\alpha)^{1/2}v_{\rm in}V_{\rm K,in}}{(2n+1)r_{\rm in}}\ \xi^{-2}, \nonumber \\
  \frac{GM}{r^2} &=& V_{\rm K,in}^2r_{\rm in}\ \xi^{-2}, \nonumber 
\end{eqnarray}
have been used. The allowed range for the parameter $n$ becomes $-1/2<n<1$ since $0<\alpha<1$ 
(in fact, there is a severer restriction $-1/4<n<1/2$, see \cite{Kab01}). 

Substituting the value of $\alpha$, we obtain the final expressions for $\tilde{v}_r$, 
$\tilde{v}_{\varphi}$, and $\tilde{\rho}$ as 
\begin{equation}
  \tilde{v}_r(\xi) = \left( \frac{2n+1}{3} \right)^{1/2}V_{\rm K,in}\ \xi^{-1}, 
  \label{eqn:vrVPF}
\end{equation}
\begin{equation}
  \tilde{v}_{\varphi}(\xi) = \left( \frac{2n+1}{3} \right)^{1/2}V_{\rm K,in}\ \xi^{-1/2}, 
\end{equation}
and 
\begin{equation}
  \tilde{\rho}(\xi) = \frac{3B_{\rm in}^2}{8\pi V_{\rm K,in}^2}\ \xi^{-(1-2n)}, 
\end{equation}
respectively. 
It is also straightforward to derive 
\begin{equation}
  \tilde{v}_{\theta}(\xi) = 2n\left( \frac{2n+1}{3} \right)^{1/2}V_{\rm K,in}\ \xi^{-1}, 
  \label{eqn:vthVPF}
\end{equation}
and 
\begin{equation}
  \tilde{T}(\xi) = \frac{V_{\rm K,in}^2}{3K}\ \xi^{-1}, 
\end{equation}
with the aid of equations (\ref{eqn:mcontS}) and (\ref{eqn:eqsS}). 
The result (\ref{eqn:vthVPF}) indicates clearly that a widely distributed wind (i.e., the vertical flow 
having the same $\xi$-dependence as $\tilde{v}_r$) appears in proportional to $n$, the degree of 
non-adiabaticity. This is the reason why we identify this wind with a thermally driven wind. It is 
interesting to see that also the vertical current is proportional to $n$ as seen in equation 
(\ref{eqn:jthVPF}). 

In this section, we have derived the VPF solution starting from the condition 
$\tilde{\mbox{\boldmath $E$}}=0$, under the restriction of the quasi-Keplerian rotation. 
The derivation is quite straightforward. 
On the other hand, the same solution was derived in Kaburaki (2000, 2001) by assuming the condition  
 \begin{equation}
     \beta \equiv \frac{1}{\tilde{b}_r}\ \frac{d}{dr}(r\tilde{v}_r\tilde{b}_{\varphi})
      \Bigm/r^2 \frac{d}{dr}\left(\frac{\tilde{v}_{\varphi}}{r}\right) = \mbox{const.}, 
 \label{eqn:beta}
 \end{equation}
instead of $\tilde{E}_r=\tilde{E}_{\theta}=0$. Since the condition $\tilde{E}_{\varphi}=0$ is 
a direct consequence of $\partial\tilde{\mbox{\boldmath$b$}}_{\rm p}/\partial t=0$ (where 
$\tilde{\mbox{\boldmath$b$}}_{\rm p}$ denotes the poloidal part of $\tilde{\mbox{\boldmath $b$}}$) 
and common to both cases, the relation (\ref{eqn:beta}) seems to be equivalent to the condition 
$\tilde{E}_r=\tilde{E}_{\theta}=0$ which is the one between the two possibilities arising from the 
condition $\partial\tilde{b}_{\varphi}/\partial t=0$. 
Thus, we have been able to replace the restriction (\ref{eqn:beta}), whose physical meaning is rather 
unclear, by the clearcut condition $\tilde{E}_r=\tilde{E}_{\theta}=0$. 

Finally, we stress here the advantages of the resistive MHD treatments over those of the ideal MHD. 
In the ideal MHD scheme, the electric field is given from the beginning by the motional field with 
negative sign, 
  \begin{equation}
    \mbox{\boldmath $E$} = - \frac{1}{c}\ \mbox{\boldmath $v$}\ \times\mbox{\boldmath $B$}, 
   \label{eqn:idMHD}
  \end{equation}
neglecting the resistive term in equation (\ref{eqn:Ohm}) as $\sigma$ is sufficiently large. 
However, this asumption poses a very stringent restriction on the possible solutions. Evidently, we cannot 
reach in the latter scheme the above discussed VPF solution since the vanishing of the electric field, 
$\mbox{\boldmath $E$}=0$, means $\mbox{\boldmath $v$}=0$ and/or $\mbox{\boldmath $B$}=0$ 
as long as the magnetic field is threading the accretion disk, which is nonsense. 

In fact, the relative size of the resistive term in Ohm's law should be determined self-consistently 
within a given problem, irrespective of the size of the conductivity. Whenever the resistive term plays 
a non-negligible contribution, we should interpret rather that the size of the electric current varies 
in proportion to the conductivity, keeping their ratio finite. In the VPF solution, we have 
  \begin{equation}
    \frac{\mbox{\boldmath $j$}}{\sigma} = \frac{1}{c}\ \mbox{\boldmath $v$}\ \times\mbox{\boldmath $B$}, 
   \label{eqn:VPF}
  \end{equation}
since $\mbox{\boldmath $E$}=0$. This means a local balance of the motional field with the 
resistive term. Therefore, the electromotive force does not remain outside the driving system, the 
accretion disk. The whole generated electromagnetic energy is dissipated locally as the Joule heating. 
We stress again that this kind of solution can never been obtained in the ideal MHD scheme. 

\section{Non-VPF Solution} 

In the previous section, we have seen that requiring a stationary magnetic field is closely related 
to the vanishment of the electric field, at least, under the assumption of quasi-Keplerian rotation. 
Since such a solution cannot explain the generation of jet-driving Poynting flux in an accretion disk, 
we have to abandon this assumption in order to proceed toward jet launching. For this reason, we have 
formulated the quasi-stationary problem in section 2. At the same time, we have to deal with the inner 
region of the accretion disks more carefully because jets are known to be launched from the vicinity of 
disk's inner edges, where $\epsilon$ approaches unity. This means that the complete neglection of 
${\cal O}(\epsilon)$ terms is insufficient. 

Thus, we now return to the original set of MHD equations obtained in section 4, without neglecting 
the left-hand side of equation (\ref{eqn:EOMrM}) and the magnetic pressure term arising from $\tilde{b}_r$ 
in equation (\ref{eqn:EOMthS}). In this new framework, we try to extend the VPF solution to be 
correct up to the first order in $\epsilon$. Of course, even the first order corrections to the VPF 
solution are not sufficient to describe exactly the regions very close to the inner edge and below it. 
Nevertheless, it is well expected that the result may give us a certain clue to the physical origin of 
jet-driving powers. 

In order to find a model case of such a solution (hereafter we call it simply the non-VPF solution), we 
put a few restrictions on the solution: 
\begin{enumerate}
  \item the non-VPF solution should coincide with the VPF solution in the limit of 
        $\xi\rightarrow\infty$ ($\mbox{i.e.,}\ \epsilon\rightarrow 0$), 
  \item as an extension of equation (\ref{eqn:alpha}), we require 
   \begin{equation}
     -\frac{r}{\tilde{\rho}}\ \frac{d}{dr}\left(\frac{\tilde{b}_{\varphi}^2}{8\pi}\right)
       \Bigm/\tilde{v}_{\varphi}^2 = \frac{\alpha}{1-\alpha}, \quad \alpha=\frac{2}{3}(1-n). 
   \label{eqn:alphaM}
   \end {equation}
\end{enumerate}

The requirement 1 is quite natural, and serves as a boundary condition for each physical quantity. In 
accordance with this requirement, we fix hereafter the position of disk's inner edge by the relation 
$\epsilon_0=1$, where $\epsilon_0$ is the value of $\epsilon$ calculated from the VPF solution, 
i.e., $\epsilon_0\equiv(\tilde{v}_r/\tilde{v}_{\varphi})^2_{\rm VPF}=\xi^{-1}$. Therefore, this place 
can be different from that calculated from $\epsilon=1$ in the non-VPF solution. 

On the other hand, the requirement 2 may seem somewhat intuitive or arbitrary. Although it is easy 
to confirm that this condition is identical with (\ref{eqn:alpha}) in the limit of $\xi\rightarrow\infty$, 
one may feel that there is no necessity to require (\ref{eqn:alphaM}) among other such possibilities. 
We agree with this assertion, but it is also true that if we admit this condition we can reach a set 
of physical quantities all of which satisfy every equation within the required accuracy. Therefore, it 
can be called a solution. Of course, however, we never insist that this is the unique solution. In this 
sense, the non-VPF solution obtained in this paper should be called a model solution. 

Eliminating the pressure $p$ in equation (\ref{eqn:EOMrS}) with the aid of equation (\ref{eqn:EOMthS}), 
we can rewrite the $r$-component of EOM as 
  \begin{eqnarray}
     \lefteqn{ \tilde{v}_r \frac{d\tilde{v}_r}{dr} 
       - \frac{\tilde{b}_r^2}{4\pi\tilde{\rho} r} 
         \left( \frac{\tilde{b}_{\theta}}{\tilde{b}_r} 
            - r\frac{d}{dr}\ln\tilde{b}_r \right) } \nonumber \\
      & & \qquad\qquad\qquad = \frac{\tilde{v}_{\varphi}^2}{r}-\frac{1}{\tilde{\rho}}
         \ \frac{d}{dr}\left(\frac{\tilde{b}_{\varphi}^2}{8\pi}\right) -\frac{GM}{r^2}, 
  \end{eqnarray}
where the terms of ${\cal O}(\epsilon)$ have been collected on the left-hand side. Further substituting 
equations (\ref{eqn:fcontS}) and (\ref{eqn:alphaM}), we finally obtain 
  \begin{eqnarray}
     \lefteqn{ \left\{ r\frac{d}{dr} \ln\tilde{v}_r 
       + \frac{r}{2D}\frac{d}{dr} \ln\left(\frac{\tilde{b}_r}{r^2} \right) \right\} 
         \tilde{v}_r^2 }\nonumber \\
      & & \qquad\qquad\qquad = \frac{1}{1-\alpha}\ 
          \left\{ \tilde{v}_{\varphi}^2 - (1-\alpha)V_{\rm K}^2 \right\}, 
  \label{eqn:Vr}
  \end{eqnarray}
where the discriminator of radially infalling matter have been defined by 
\begin{equation}
  D \equiv \frac{\tilde{v}_r^2}{V_{\rm A}^2},  
  \label{eqn:defD}
\end{equation}
with $V_{\rm A} \equiv \tilde{b}_r/(4\pi\tilde{\rho})^{1/2}$ being the Alfv\'{e}n velocity. 

Since the factor $\tilde{v}_r^2$ on the left-hand side of equation (\ref{eqn:Vr}) is already a 
quantity of ${\cal O}(\epsilon)$ by definition, it is sufficient to evaluate the factor in the curly 
brackets by using the lowest order solution, i.e., the VPF solution. 
The values of the related quantities evaluated in terms of the VPF solution are 
\begin{equation}
  [D]_{\rm VPF} = \frac{2n+1}{2}, 
\end{equation}
\begin{equation}
  \left[ r\frac{d}{dr} \ln\tilde{v}_r 
   + \frac{r}{2D}\frac{d}{dr} \ln\left(\frac{\tilde{b}_r}{r^2} \right) \right]_{\rm VPF} 
   = -\frac{2n+9}{2n+1}. 
\end{equation}
It is evident from the latter result that the left-hand side of equation (\ref{eqn:Vr}) is negative, 
and hence, it follows that $\tilde{v}_{\varphi}<(1-\alpha)^{1/2}V_{\rm K}$. This means that, owing to the 
inclusion of the inertial and magnetic draggings on the left-hand side, the centrifugal force required 
to balance the gravitational attraction can be reduced from the VPF case (especially in the inner regions). 
In the limit of $\xi\rightarrow\infty$, we have $\tilde{v}_{\varphi} \rightarrow (1-\alpha)^{1/2}V_{\rm K}$ 
from equation (\ref{eqn:Vr}), neglecting the terms of ${\cal O}\sim(\epsilon)$ on the left-hand side. 

Although these constraints are not adequate to determine the functional form of $\tilde{v}_{\varphi}$ 
uniquely, here we anticipate its from as 
\begin{equation}
  \tilde{v}_{\varphi}(\xi) 
    = \left(\frac{2n+1}{3}\right)^{1/2}V_{\rm K,in}\ \xi^{-1/2}e^{-A\xi^{-1}}, 
\end{equation}
in order to obtain an illustrative model solution of the non-VPF version. 
Substituting this expression into equation (\ref{eqn:Vr}), we obtain the expression for $\tilde{v}_r^2$ 
containing a parameter $A$. The value of $A$ can be determined by the condition, 
$\tilde{v}_r \rightarrow [\tilde{v}_r]_{\rm VPF}$ as $\xi\rightarrow \infty$. The results are 
\begin{eqnarray}
   \tilde{v}_r(\xi) 
    &=& \left(\frac{2n+1}{3}\right)^{1/2}V_{\rm K,in}\ \xi^{-1/2}
      \left(\frac{1-e^{-2A\xi^{-1}}}{2A}\right)^{1/2}  \nonumber\\ 
    &\approx& \left(\frac{2n+1}{3}\right)^{1/2}V_{\rm K,in}\ \xi^{-1} 
      \quad (\mbox{for }\ \xi\gg 1), 
   \label{eqn:VrC}
\end{eqnarray}
where 
\begin{equation}
  A = \frac{2n+9}{12}. 
\end{equation}
The value $1-\alpha=(2n+1)/3$ has been substituted in the above derivation. 

The last expression for $\tilde{v}_r(\xi)$ is obtained by expanding the exponential factor in a power 
series of $\epsilon_0=\xi^{-1}$ and keeping only the terms up to the first order. In the accretion disk 
(where $\xi\geq 1$), the approximation to this order is sufficient for the present purpose. Namely, the 
inclusion of the first order corrections in $\epsilon$ has not altered the form of $\tilde{v}_r$ from 
that in the VPF solution, equation (\ref{eqn:vrVPF}). It should be noted, however, that the former has 
been derived from the $r$-component of EOM instead of Ohm's law. The ratio of the velocity components 
then becomes 
\begin{equation}
  \frac{\tilde{v}_{\varphi}}{\tilde{v}_r} 
    =  e^{-A\xi^{-1}} \left(\frac{1-e^{-2A\xi^{-1}}}{2A}\right)^{-1/2} 
    \approx \xi^{1/2}e^{-A\xi^{-1}}. 
  \label{eqn:Vratio}
\end{equation}
This means that $\epsilon\equiv (\tilde{v}_r/\tilde{v}_{\varphi})^2\approx\xi^{-1}\exp(2A\xi^{-1})$. 
Thus, the actual value of $\epsilon$ deviates largely from $\epsilon_0$ near and below the inner edge 
($\xi=1$). 

Next, we rewrite the $\varphi$-component of EOM, (\ref{eqn:EOMphS}), with the aid of equation 
(\ref{eqn:defD}). Eliminating $1/4\pi\tilde{\rho}$ from the former, we obtain first  
\begin{displaymath}
  \frac{\tilde{b}_{\theta}\tilde{b}_{\varphi}}{\tilde{b}_r^2} 
  = D\ \frac{d\tilde{l}/dr}{\tilde{v}_r}, \quad \tilde{l}\equiv r\tilde{v}_{\varphi}, 
\end{displaymath}
and further using equation (\ref{eqn:fcontS}) and the definition of $\Re_t$, 
\begin{equation}
  \Re_t = 2D \left\{r\frac{d}{dr}\ln\tilde{l}\Bigm/r\frac{d}{dr}\ln(r^2\tilde{b}_r) \right\} 
          \ \frac{\tilde{v}_{\varphi}}{\tilde{v}_r}.  
  \label{eqn:EQam}
\end{equation}
Similarly, the extended RIAF condition (\ref{eqn:alphaM}) can be rewritten in the from 
\begin{equation}
  -r\frac{d}{dr}\ln\tilde{b}_{\varphi} 
   = \frac{2(1-n)}{2n+1}\ D \left( \frac{\tilde{v}_{\varphi}}{\Re_t\tilde{v}_r} \right)^2.  
\end{equation}
Since the ratio $\tilde{v}_{\varphi}/\tilde{v}_r$ in our model case is approximated by the last expression 
in (\ref{eqn:Vratio}), the above equation reads 
\begin{equation}
  -r\frac{d}{dr}\ln\tilde{b}_{\varphi} 
   = \frac{1-n}{2n+1}\ \xi\ e^{-2A\xi^{-1}}\ \frac{2D}{\Re_t^2}. 
  \label{eqn:xRIAF}
\end{equation}

As a trial function for $\tilde{b}_r$, we chose here 
\begin{equation}
  \tilde{b}_r(\xi) = B_{\rm in}\ \xi^{-(3/2-n)}e^{-(2n+1)A\xi^{-1}}. 
\end{equation}
The exponential factor added to the corresponding VPF solution causes a rapid decrease of this 
component at radii smaller than the disk inner edge. This seems very reasonable since it should vanish 
on the polar axes owing to the symmetry requirement. The above $\tilde{b}_r$ results in 
\begin{eqnarray}
  \lefteqn{ \tilde{b}_{\theta}(\xi) = \frac{2n+1}{4}B_{\rm in}\times }\nonumber \\
    & & \qquad\qquad \times(1+2A\xi^{-1})\ \xi^{-(3/2-n)}e^{-(2n+1)A\xi^{-1}}, 
\end{eqnarray}
from the flux conservation equation (\ref{eqn:fcontS}). 
Then, further noting that $rd\ln\tilde{l}/dr=(1+2A\xi^{-1})/2$, we obtain 
\begin{equation}
  \Re_t(\xi) = \frac{2D}{2n+1}\ \xi^{1/2}e^{-A\xi^{-1}} 
  \label{eqn:Ret}
\end{equation}
from equation (\ref{eqn:EQam}).

It is very interesting to examine the possibility in which 
\begin{equation}
  D(\xi) = \frac{2n+1}{2}\ e^{A\xi^{-1}}. 
  \label{eqn:accel}
\end{equation}
Of course, this satisfies the requirement $D\rightarrow (2n+1)/2$ since $e^{A\xi^{-1}}\rightarrow 1$ 
in the limit of $\xi\rightarrow \infty$. The discriminator selected above corresponds to a 
trans-critically accelerated infall for any allowed value of $n$ ($-1/4<n<1/2$). Such an infall is 
analogous to that in Bondi's theory of spherical accretion, although no special affairs appear at the 
`critical point' (i.e., at the point where $\tilde{v}_r=V_{\rm A}$ or $D=1$) in our problem. 
If the discriminant is given by the above formula, then equations (\ref{eqn:Ret}) and 
(\ref{eqn:xRIAF}) reduce, respectively, to 
\begin{equation}
  \Re_t(\xi) = \xi^{1/2}, 
  \label{eqn:RetF}
\end{equation}
and 
\begin{equation}
  -r\frac{d}{dr}\ln\tilde{b}_{\varphi} = (1-n)\ e^{-A\xi^{-1}}. 
  \label{eqn:fRIAF}
\end{equation}
The form of $\Re_t$ is thus identical to that in the VPF solution. 

We can easily confirm that the functional form 
\begin{equation}
  \tilde{b}_{\varphi}(\xi) = B_{\rm in}\ \xi^{-(1-n)}e^{-(1-n)A\xi^{-1}} 
\end{equation}
satisfies equation (\ref{eqn:fRIAF}) within the required accuracy, since 
$1-A\xi^{-1}\approx e^{-A\xi^{-1}}$ up to $\xi^{-1} (= \epsilon_0)$. In order for this 
$\tilde{b}_{\varphi}$ to be a non-VPF solution, it should be consistent with equation (\ref{eqn:RetF}). 
Since we have 
\begin{displaymath}
  \Re_t \equiv \frac{\tilde{b}_{\varphi}}{\tilde{b}_r} = \xi^{1/2}e^{3nA\xi^{-1}}, 
\end{displaymath}
it turns out that the three components of the magnetic field introduced in this section and the 
postulate (\ref{eqn:accel}) are consistent with the set of quasi-stationary MHD equations if and only 
if $n=0$, i.e., the accretion flow is adiabatic. 
Thus, the value of the parameter $n$, which cannot be determined in the framework of the lowest order 
approximation, is restricted to the special case of $n=0$ in our improved treatment, without referring 
to the energy equation. This fact suggests that the realization of the present, non-VPR model solution 
in actual situations sensitively depends on the affairs of energy transport which can be realized in 
the accretion flows. 

Thus, the final forms of the magnetic field components in our model solution (in which $n=0$) become 
\begin{equation}
  \tilde{b}_r(\xi) = B_{\rm in}\ \xi^{-3/2}e^{-A\xi^{-1}}, 
\end{equation}
\begin{equation}
  \tilde{b}_{\theta}(\xi) = \frac{1}{4}B_{\rm in}\ (1+2A\xi^{-1})\xi^{-3/2}e^{-A\xi^{-1}}, 
\end{equation}
\begin{equation}
  \tilde{b}_{\varphi}(\xi) = B_{\rm in}\ \xi^{-1}e^{-A\xi^{-1}}, 
\end{equation}
with 
\begin{equation}
  A = \frac{3}{4}. 
\end{equation}
The discriminator of the radial flow is specified as 
\begin{equation}
  D(\xi) = \frac{1}{2}\ e^{A\xi^{-1}}. 
\end{equation}

The derivation of remaining quantities are straightforward. We obtain 
\begin{equation}
  \tilde{\rho}(\xi) = \frac{3B_{\rm in}^2}{8\pi V_{\rm K,in}^2}\ \xi^{-1}e^{-A\xi^{-1}}
\end{equation}
from equation (\ref{eqn:EOMphS}), 
\begin{equation}
  \tilde{p}(\xi) = \frac{B_{\rm in}^2}{8\pi}\ (1+\xi^{-1})\ \xi^{-2}e^{-2A\xi^{-1}} 
\end{equation}
from equation (\ref{eqn:EOMthS}), and 
\begin{equation}
  \tilde{T}(\xi) = \frac{V_{\rm K,in}^2}{3K}\ (1+\xi^{-1})\ \xi^{-1}e^{-A\xi^{-1}} 
\end{equation}
from equation (\ref{eqn:eqsS}). 
Here, the deep suppression of the temperature from the corresponding VPF value near the inner edge 
should be noted, since it has already been expected from spectral fittings for some LLAGN based on the 
VPF solution (e.g., \cite{YKK02, KNTW10}). The suppressions in the density and pressure are consistent 
with the presence of outflows from the accretion disk in the forms of jets.  

The final (i.e., when $n=0$) forms of $\tilde{v}_r$ and $\tilde{v}_{\varphi}$ are 
\begin{equation}
  \tilde{v}_r(\xi) = \frac{V_{\rm K,in}}{\sqrt{3}}\ \xi^{-1}, 
\end{equation}
\begin{equation}
  \tilde{v}_{\varphi}(\xi) = \frac{V_{\rm K,in}}{\sqrt{3}}\ \xi^{-1/2}e^{-A\xi^{-1}}, 
\end{equation}
and equation (\ref{eqn:mcontS}) yields 
\begin{equation}
  \tilde{v}_{\theta}(\xi) = \frac{AV_{\rm K,in}}{\sqrt{3}}\ \xi^{-2}. 
 \label{eqn:vphF}
\end{equation}
As confirmed from the angular dependence in (\ref{eqn:vth}), $v_{\theta}(\xi,\ \eta)$ resulting 
from the above $\tilde{v}_{\theta}\ (>0)$ describes an outflow from the accretion disk. 
Different from the case of the VPF solution, $\tilde{v}_{\theta}$ here exists even in the case of 
$n=0$. Corresponding to this fact, the right-hand side of equation (\ref{eqn:vphF}) contains the 
constant $A$ instead of $n$ (compare with $\tilde{v}_{\theta}(\xi)$ in the VPF solution). The functional 
form of this outflow is also a power law but steeper than the radial infall of the present solution 
(see figure 1) and the vertical outflow in the VPF solution. These are the reasons why we expect 
that the vertical outflow in the non-VPF solution corresponds to the base of an electromagnetically 
driven jet, distinguished from the thermally driven wind in the VPF solution. 

\section{Poynting Flux, Time Variability and Bernoulli Sum}

The components of the current density in our non-VPF model solution are derived from Amp\`{e}re's 
law as 
\begin{equation}
  \tilde{j}_r(\xi) = \frac{cB_{\rm in}}{4\pi r_{\rm in}}\ \xi^{-2}e^{-A\xi^{-1}}, 
\end{equation}
\begin{equation}
  \tilde{j}_{\theta}(\xi) = \frac{cAB_{\rm in}}{4\pi r_{\rm in}}\ \xi^{-2}e^{-A\xi^{-1}}, 
\end{equation}
\begin{equation}
  \tilde{j}_{\varphi}(\xi) = \frac{cB_{\rm in}}{4\pi r_{\rm in}}\ \xi^{-5/2}e^{-A\xi^{-1}}. 
\end{equation}
Similarly to the case of $\tilde{v}_{\theta}$, $\tilde{j}_{\theta}$ contains the constant $A$ instead 
of $n$ in the corresponding VPF versions. The presence of this component even in the case 
of $n=0$ would be important from the viewpoint of current closure. 

\begin{figure}


 \begin{center}
  \FigureFile(85mm, 95mm){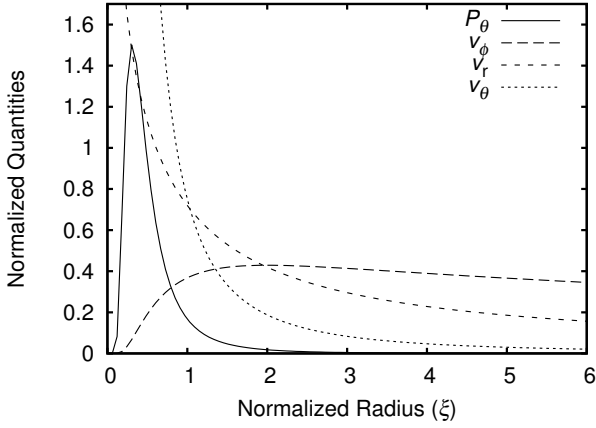}
 \end{center}
 \caption{Radial profiles of the $\theta$-component of the Poynting flux ($\tilde{P}_{\theta}$) and 
the velocity components of the accretion flow ($\tilde{v}_r$, $\tilde{v}_{\theta}$, $\tilde{v}_{\varphi}$). 
They are normalized to $AV_{\rm K,in}B^2_{\rm in}/4\sqrt{3}\pi$ and $V_{\rm K,in}/\sqrt{3}$, respectively. 
The profiles are written based on the functional forms obtained in the text and shown extended to the 
inner region ($\xi<1$) where the accuracy of the solution may be poor ($v_r$ is based on the first 
expression in equation (\ref{eqn:VrC})). The downward deviation of the rotational velocity ($v_{\varphi}$) 
from the Kepler curve becomes appreciable below $\xi\simeq3$ and the vertical velocity exceeds 
the radial one only below $\xi\simeq1$. The vertical Poynting flux has a sharp peak near the disk inner 
edge ($\epsilon_0=\xi^{-1}=1$). }
\end{figure}

In order to calculate the full components of the electric field from Ohm's law, we need the value 
of $\Re_{\rm p}$. This can be evaluated from the definitions of $v_{\rm in}$ in (\ref{eqn:vrm}) 
and of $\Re_{\rm p}$ in (\ref{eqn:defRe}) as 
\begin{equation}
  \Re_{\rm p}(\xi) = \frac{4}{2n+1}\ \xi\frac{\tilde{v}_r}{v_{\rm in}}
      \frac{\tilde{b}_{\theta}}{\tilde{b}_r} = 1+2A\xi^{-1}, 
\end{equation}
where the value of $v_{\rm in}$ has been fixed by the condition $\Re_{\rm in}=1$. Using this result, 
we obtain 
\begin{equation}
  \tilde{E}_r(\xi) = \frac{AV_{\rm K,in}B_{\rm in}}{4\sqrt{3}c}\ \xi^{-5/2}e^{-A\xi^{-1}}, 
\end{equation}
\begin{eqnarray}
  \tilde{E}_{\theta}(\xi) &=& -\frac{V_{\rm K,in}B_{\rm in}}{\sqrt{3}c}
    \ \xi^{-2}e^{-A\xi^{-1}}\left( 1- e^{-A\xi^{-1}}\right) \nonumber\\
     &\approx& -\frac{AV_{\rm K,in}B_{\rm in}}{\sqrt{3}c}\ \xi^{-3}e^{-A\xi^{-1}}, 
\end{eqnarray}
\begin{equation}
  \tilde{E}_{\varphi}(\xi) = -\frac{AV_{\rm K,in}B_{\rm in}}{2\sqrt{3}c}\ \xi^{-7/2}e^{-A\xi^{-1}}. 
\end{equation}

The calculation of the Poynting flux is straightforward, and the resulting components are  
\begin{equation}
  \tilde{P}_r(\xi) = -\frac{AV_{\rm K,in}B_{\rm in}^2}{4\sqrt{3}\pi}\ \xi^{-4}e^{-2A\xi^{-1}}, 
\end{equation}
\begin{equation}
  \tilde{P}_{\theta}(\xi) = \frac{AV_{\rm K,in}B_{\rm in}^2}{16\sqrt{3}\pi}\ 
     \xi^{-7/2}\left(1+2\xi^{-3/2}\right)\ e^{-2A\xi^{-1}}, 
\end{equation}
\begin{equation}
  \tilde{P}_{\varphi}(\xi) = -\frac{AV_{\rm K,in}B_{\rm in}^2}{4\sqrt{3}\pi}\ \xi^{-9/2}e^{-2A\xi^{-1}}. 
\end{equation}
The existence of a steep negative power-law dependence in $\xi$ and  that of an exponential factor in 
each of these components clearly indicates a localized nature of the Poynting flux to the region near 
or even within the disk inner edge (see figure 1). Since we have 
\begin{equation}
  {\bf \nabla}\cdot\mbox{\boldmath $P$} = \frac{\tilde{P}_{\theta}(\xi)}{r}\ {\rm sech}^4\eta, 
\end{equation}
the positivity of $\tilde{P}_{\theta}$ guarantees that the accretion disk is an electric power source 
and is driving outside system, most probably, a bipolar-jet system. Although the structures of the 
resistive MHD jets at rather remote regions from their footpoints have been discussed in our previous 
work (\cite{Kab09}), further detailed investigations are required to fully describe the jet 
launching processes. 

In our quasi-stationary treatment, the time variabilities in the magnetic field can be calculated 
according to equations (\ref{eqn:FdrS}) $\sim$ (\ref{eqn:FdphS}) as 
\begin{equation}
   \frac{1}{c}\frac{\partial \tilde{b}_r}{\partial t} 
    = \frac{AV_{\rm K,in}B_{\rm in}}{2\sqrt{3}\ cr_0}\ \xi^{-9/2}\ e^{-A\xi^{-1}},
\end{equation}
\begin{equation}
   \frac{1}{c}\frac{\partial \tilde{b}_{\theta}}{\partial t} 
    = -\frac{5AV_{\rm K,in}B_{\rm in}}{4\sqrt{3}\ cr_0}
      \left( 1-\frac{2}{5}A\xi^{-1} \right)\ \xi^{-9/2}\ e^{-A\xi^{-1}},
\end{equation}
\begin{eqnarray}
   \lefteqn{ \frac{1}{c}\frac{\partial \tilde{b}_{\varphi}}{\partial t} 
    = \frac{AV_{\rm K,in}B_{\rm in}}{2\sqrt{3}\ cr_0}\times } \nonumber \\
     & & \qquad\quad \times\left( 1-4\xi^{-1/2}+2A\xi^{-3/2} \right)\ \xi^{-5/2}\ e^{-A\xi^{-1}}. 
\end{eqnarray}
As is evident from the presences of the exponential and steep power-law factors also in these 
expressions, the time variations become noticeable only in a narrow region near the disk inner edge. 
The characteristic time scale of these changes is that of the infall according to the requirement 3 in 
section 2, and is very short in this inner region supposed to be the base of a jet. This fact is very 
interesting because such a rapid time variation in the magnetic field may well explain the presence of 
rapid phenomina such as flares or bursts often observed at the roots of various jets. For example, 
we can imagine that a large accumulation of the magnetic field would cause an impulsive acceleration 
of fast jet. 

The Bernoulli sum is the sum of energy carried along with a fluid element per unit mass, and is defined 
here as  
\begin{equation}
  \tilde{B}_{\rm e}(\xi) \equiv \frac{1}{2}\left( \tilde{v}_r^2 + \tilde{v}_{\varphi}^2 \right) 
   + \frac{5}{2}\ \frac{\tilde{p}}{\tilde{\rho}} - \frac{GM}{r},  
\end{equation}
where the second term on the right-hand side is the enthalpy of an ideal gas. The kinetic energy 
associated with $\tilde{v}_{\theta}$ has been neglected because it is a small quantity of 
${\cal O}(\Delta^2)$. Here, we pay attention only to $\tilde{B}_{\rm e}(\xi)$ although the $\eta$-dependence 
of each term is different from each other. This is because our primary interest is on the midplane of 
an accretion disk around which the majority of matter is concentrated. The lack of the electromagnetic 
field energy in the sum is due to the fact that the field is distributed in space and is not carried by 
the fluid motion in the quasi-stationary situation (in contrast to the ideal MHD case). 

For a discussion of overall behaviors of the Bernoulli sum, we have to extrapolate the results of our 
non-VPF model solution beyond the inner edge toward the center (i.e., $\xi\lesssim 1$). For this purpose, 
the original expression for $\tilde{v}_r$ given in (\ref{eqn:VrC}) is suited. The final result calculated 
in terms of the non-VPF solution is 
\begin{eqnarray}
 \lefteqn{ \tilde{B}_{\rm e}(\xi) = -\frac{8}{9}\ V_{\rm K,in}^2\xi^{-1} \times } \nonumber \\
  & & \qquad\qquad \times \left\{ 1-\left(1+\frac{15}{16}\ \xi^{-1}\right)\ e^{-(3/2)\xi^{-1}} \right\}, 
\end{eqnarray}
where the value $A=3/4$ has been substituted. The asymptotic values for large and small $\xi$ are 
\begin{equation}
 \tilde{B}_{\rm e}(\xi) \approx \left\{
   \begin{array}{rl}
     -\displaystyle{\frac{1}{2}}\ V_{\rm K,in}^2 \xi^{-2},&\quad \mbox{as }\ \xi \rightarrow \infty, \\
     \noalign{\vskip 0.2cm}
     -\displaystyle{\frac{8}{9}}\ V_{\rm K,in}^2 \xi^{-2},&\quad \mbox{as }\ \xi \rightarrow 0.  
   \end{array} \right. 
\end{equation}
In the VPF solution, we have $\tilde{B}_{\rm e}(\xi)=0$ within the accretion disk ($\xi\geq 1$) when 
$n=0$ (\authorcite{Kab00}\yearcite{Kab00,Kab01}). The above expression for the limit of 
$\xi\rightarrow\infty$ includes the first order correction in $\xi^{-1}$ to that result. In both large 
and small radius limits, $\tilde{B}_{\rm e}(\xi)$ behaves like $\xi^{-2}$ having negative and increasingly 
large absolute values toward the center, which reflects the nature of infalling matter. 

\section{Summary and Conclusion}

The resistive RIAF (radiatively inefficient accretion flow) model has been very successful as a model 
for the inefficiently radiating accretion disks threaded by an external, ordered magnetic field. 
It provides a firm basis to discuss various physical processes taking place in such accretion flows. 
It can also well reproduce the main properties of observed broad-band spectral energy distributions 
emanating from the nuclei of nearby normal galaxies and some types of active galactic nuclei, which are 
all believed to be accreting at a very small fraction of each Eddington accretion rate. The present 
status of this model, however, is unsatisfactory in that it cannot explain the electrodynamic formation 
and acceleration of bipolar jets which are frequently observed associated with such accretion flows. 
This is because the Poynting flux which is necessary to drive outside systems (especially a jet) 
vanishes identically. 

In order to overcome this defect in the current version of our model, we have carefully 
reformulated in the present paper the problem we have really to deal with. The new scheme 
allows a quasi-stationary change of the magnetic field, and hence, the presence of a non-irrotational 
electric field. 
Another imrpovement in the new scheme is the restoration of the terms of ${\cal O}(\epsilon)$, 
where $\epsilon\equiv(v_r/v_{\varphi})^2$, which have been neglected as small. This is to keep the 
problem valid even in the inner regions where $\epsilon\sim 1$. 
The main restored terms are the inertial force of the infalling matter and the dragging force 
due to the poloidal magnetic field. Both these terms appear in the radial component of EOM, and 
reduces the rotational velocity of an accretion disk near its inner edge. 

As an example of the solutions to this modified problem, the non-VPF (vanishing Poynting-flux) 
solution which is accurate up tp the first order in $\epsilon$, has been derived under some plausible 
constraints. Although this solution has many essential advantages over the previous version (the 
VPF solution), it is still an approximate solution in which the accuracy becomes poor near the disk 
inner edge. Therefore, one should be well aware of the possible inaccuracies when he applies the obtained 
results to such regions. 

The obtained solution has a non-vanishing electric field and a diverging Poynting flux, the latter 
of which suggests that the accretion disk is electrodynamically driving some outside systems. 
Interestingly, the model holds only when the RIAF is, in particular, an ADAF (advection dominated 
accretion flow), that is, when the accretion flow is adiabatic and hence no thermally driven wind from 
the disk surfaces exists. The predicted vertical components of the Poynting flux and the outflow at 
the disk surface are much localized to the region very close to the disk inner edge, strongly suggesting 
that the resistive RIAF disk has an ability to drive electromagmetic jet. 
This result seems to be quite consistent with the observational fact that strong jets are often associated 
with the nuclei accreting at very small fractions of their Eddington rates. In such rarefied 
accretion disks, the radiation losses and energy exchange among accreting matter cannot occur efficiently. 

The obtained model also suggests the presence of rapid changes in the magnetic field localized to 
the inner acceleration region. This fact may explain the origin of some violent phenomena such as the 
flares and/or impulsive acceleration of strong jets. 

Thus, our resistive RIAF (or ADAF) model has been greatly improved to be able to admit the formation 
of electrodynamic jets in the region very close to the inner edge of an accretion disk. 
However, for the detailed understanding of concrete mechanisms of the jet launching further 
investigations are needed.


\appendix
\section{Expressions in Spherical Polar Coordinates}

The expressions in spherical polar coordinates of the basic equations for the quasi-stationary 
problems are summarized below. The assumption of axisymmetry has already been taken into account. 
\begin{itemize}
 \item magnetic flux conservation:
  \begin{equation}
    \frac{1}{r^2}\frac{\partial}{\partial r}(r^2 b_r) 
    + \frac{1}{r\sin\theta}\ \frac{\partial}{\partial \theta}(\sin\theta\ b_{\theta}) = 0.
  \label{eqn:fcont}
  \end{equation}
 \item Amp\`{e}re's law: 
  \begin{equation}
    \frac{4\pi}{c} j_r 
       = \frac{1}{r\sin\theta}\frac{\partial}{\partial\theta} (\sin\theta\ b_{\varphi}), 
  \end{equation}
  \begin{equation}
    \frac{4\pi}{c} j_{\theta}
       = - \frac{1}{r}\frac{\partial}{\partial r} (r b_{\varphi}), 
  \end{equation}
  \begin{equation}
    \frac{4\pi}{c} j_{\varphi}
       = \frac{1}{r}\left\{\frac{\partial}{\partial r} (r b_{\theta}) 
       - \frac{\partial b_r}{\partial\theta}\right\}, 
  \end{equation}
 \item Ohm's law: 
  \begin{equation}
    E_r = \frac{j_r}{\sigma} - \frac{1}{c}(v_{\theta}b_{\varphi} - v_{\varphi}b_{\theta}), 
  \end{equation}
  \begin{equation}
    E_{\theta} = \frac{j_{\theta}}{\sigma} - \frac{1}{c}(v_{\varphi}b_r - v_rb_{\varphi}), 
  \end{equation}
  \begin{equation}
    E_{\varphi} = \frac{j_{\varphi}}{\sigma} - \frac{1}{c}(v_r b_{\theta} - v_{\theta}b_r), 
  \end{equation}
 \item mass continuity:
  \begin{equation}
    \frac{1}{r^2}\frac{\partial}{\partial r}(r^2\rho v_r) 
    + \frac{1}{r\sin\theta}\ \frac{\partial}{\partial \theta}(\sin\theta\ \rho v_{\theta}) = 0,  
  \label{eqn:mcont}
  \end{equation}
 \item equation of motion:
  \begin{eqnarray}
      \lefteqn{ \left( v_r \frac{\partial}{\partial r} 
         + \frac{v_{\theta}}{r} \frac{\partial}{\partial \theta} \right)v_r 
         - \frac{1}{r}(v_{\theta}^2+v_{\varphi}^2) }\nonumber \\
       & & \quad = -\frac{1}{\rho}\ \frac{\partial p}{\partial r} -\frac{GM}{r^2} 
         - \frac{1}{4\pi\rho r} \times \nonumber \\ 
        & & \qquad \times
         \left[ b_{\theta} \left\{\frac{\partial}{\partial r}(r b_{\theta}) 
         - \frac{\partial b_r}{\partial \theta} \right\} 
         + b_{\varphi} \frac{\partial}{\partial r}(rb_{\varphi}) \right],
  \label{eqn:EOMr}
  \end{eqnarray}
  \begin{eqnarray}
      \lefteqn{ \left( v_r \frac{\partial}{\partial r} 
         + \frac{v_{\theta}}{r} \frac{\partial}{\partial \theta} \right)v_{\theta} 
         + \frac{1}{r}(v_r v_{\theta} - v_{\varphi}^2\cot\theta) }\nonumber \\
      & & \quad = -\frac{1}{\rho r}\ \frac{\partial p}{\partial \theta} 
        -\frac{1}{4\pi\rho r}\left[ \frac{b_{\varphi}}{\sin\theta} 
        \frac{\partial}{\partial \theta}(\sin\theta\ b_{\varphi}) \right. \nonumber \\
      & & \qquad\qquad\qquad\qquad \left. -b_r\left\{ \frac{\partial}{\partial r}(r b_{\theta}) 
        - \frac{\partial b_r}{\partial\theta} \right\}\right],  
  \label{eqn:EOMth}
  \end{eqnarray}
  \begin{eqnarray}
      \lefteqn{ \left( v_r \frac{\partial}{\partial r} 
         + \frac{v_{\theta}}{r} \frac{\partial}{\partial \theta} \right)v_{\varphi} 
         + \frac{v_{\varphi}}{r}(v_r+v_{\theta}\cot\theta) }\nonumber \\
       & & \quad = \frac{1}{4\pi\rho r} \left[ b_r \frac{\partial}{\partial r}(r b_{\varphi}) 
       + \frac{b_{\theta}}{\sin \theta} 
         \frac{\partial}{\partial \theta}(\sin\theta\ b_{\varphi}) \right],
  \label{eqn:EOMph}
  \end{eqnarray}
 \item equation of state: 
  \begin{equation}
     p = K\rho T, 
  \end{equation}
 \item Faraday's law: 
  \begin{equation}
   \frac{1}{c}\ \frac{\partial b_r}{\partial t} 
   = -\frac{1}{r\sin\theta}\ \frac{\partial}{\partial\theta}(\sin\theta\ E_{\varphi}), 
  \end{equation}
  \begin{equation}
   \frac{1}{c}\ \frac{\partial b_{\theta}}{\partial t} 
   = \frac{1}{r}\ \frac{\partial}{\partial r}(rE_{\varphi}), 
  \end{equation}
  \begin{equation}
   \frac{1}{c}\ \frac{\partial b_{\varphi}}{\partial t} 
   = -\frac{1}{r}\ \left\{ \frac{\partial}{\partial r}(rE_{\theta}) 
      -\frac{\partial E_r}{\partial\theta} \right\}, 
  \end{equation}
 \item charge density: 
  \begin{eqnarray}
   \lefteqn{ q = \frac{1}{4\pi} \left\{ \frac{1}{r^2}\frac{\partial}{\partial r}(r^2E_r) 
       \right. }\nonumber\\
   & & \qquad\qquad\qquad \left.+ \frac{1}{r\sin\theta}
       \frac{\partial}{\partial\theta}(\sin\theta\ E_{\theta}) \right\}. 
  \label{eqn:chgA}
  \end{eqnarray}
\end{itemize}

\end{document}